\documentclass[conference]{IEEEtran}
\IEEEoverridecommandlockouts
\usepackage{cite}
\usepackage{amsmath,amssymb,amsfonts}
\usepackage{algorithmic}
\usepackage{graphicx}
\usepackage{textcomp}
\usepackage{xcolor}
\usepackage{stmaryrd}

\newcommand{\flow}{\mathsf{flow}}
\newcommand{\jump}{\mathsf{jump}}
\newcommand{\inv}{\mathsf{inv}}
\newcommand{\init}{\mathsf{init}}
\newcommand{\guard}{\mathsf{guard}}
\newcommand{\reset}{\mathsf{reset}}
\newcommand{\reach}{\mathsf{Reach}}
\newcommand{\unsafe}{\mathsf{unsafe}}

\newcommand{\lrf}{\mathcal{L}_{\mathbb{R}_{\mathcal{F}}}}
\newtheorem{definition}{Definition}
\newtheorem{theorem}{Theorem}

\newcommand{\citep}{\cite}

\newcommand{\R}{\mathbb{R}}

\newcommand{\tildex}{\raise.17ex\hbox{$\scriptstyle\mathtt{\sim}$}}

\def\BibTeX{{\rm B\kern-.05em{\sc i\kern-.025em b}\kern-.08em
    T\kern-.1667em\lower.7ex\hbox{E}\kern-.125emX}}
\begin{document}
\bstctlcite{IEEEexample:BSTcontrol}

\title{A Model Checking-based Analysis Framework for Systems Biology Models
}

\author{\IEEEauthorblockN{Bing Liu}
\IEEEauthorblockA{\textit{Department of Computational \& Systems Biology} \\
\textit{School of Medicine, University of Pittsburgh}\\
Pittsburgh, PA 15237, U.S.A. \\
liubing@pitt.edu}
}
\maketitle

\begin{abstract}
Biological systems are often modeled as a system of ordinary differential equations (ODEs) with time-invariant parameters. However, cell signaling events or pharmacological interventions may alter the cellular state and induce multi-mode dynamics of the system. Such systems are naturally modeled as hybrid automata, which possess multiple operational modes with specific nonlinear dynamics in each mode. In this paper we introduce a model checking-enabled framework than can model and analyze both single- and multi-mode biological systems. We tackle the central problem in systems biology--identify parameter values such that a model satisfies desired behaviors--using bounded model checking. We resort to the delta-decision procedures to solve satisfiability modulo theories (SMT) problems and sidestep undecidability of reachability problems. Our framework enables several analysis tasks including model calibration and falsification, therapeutic strategy identification, and Lyapunov stability analysis. We demonstrate the applicablitliy of these methods using case studies of prostate cancer progression, cardiac cell action potential and radiation diseases.
\end{abstract}

\begin{IEEEkeywords}
systems biology, model checking, hybrid systems, delta-decision, parameter synthesis 
\end{IEEEkeywords}

\section{Introduction}

Biomolecules interact with each other and form large and complicated networks.  A systems-wide view of functional regulation in the context of biochemical networks is required for contemporary drug discovery and systems pharmacology to identify new drug targets, understand drug adverse effects, and design therapeutic strategies~\cite{Sorger2012}. It is commonly recognized that systems modeling will play a crucial role in this endeavor. 

A standard approach of modeling the dynamics of a biochemical network is through a system of ordinary differential equations (ODEs)~\cite{Liu2012}. However, model parameters including kinetic rate constants and initial concentrations of molecular species are often unknown and they will have to be estimated using limited and noisy experimental data. Hence constructing and calibrating ODE models of realistic biochemical networks is a challenging problem. Further, ODE models that were developed for simulating single-cell behaviors were often calibrated and validated using population-based data. This constitutes a significant additional challenge.	

To address these challenges, we have developed several {\em model checking}-enabled analysis techniques for ODE models. For example, by reducing the ODE dynamics as a dynamic Bayesian network (DBN)~\cite{liu2009probabilistic,liu2011probabilistic,Liu2012b,Hagiescu2013}, one can efficiently perform parameter estimation and {\em probabilistic model checking} analysis by exploiting Bayesian inferencing~\cite{palaniappan2012hybrid,palaniappan2016look}. This framework has been applied to study complement system~\cite{Liu2011} and apoptosis pathway~\cite{palaniappan2017abstracting}. 

\begin{figure*}[t]
\centerline{\includegraphics[width=7in]{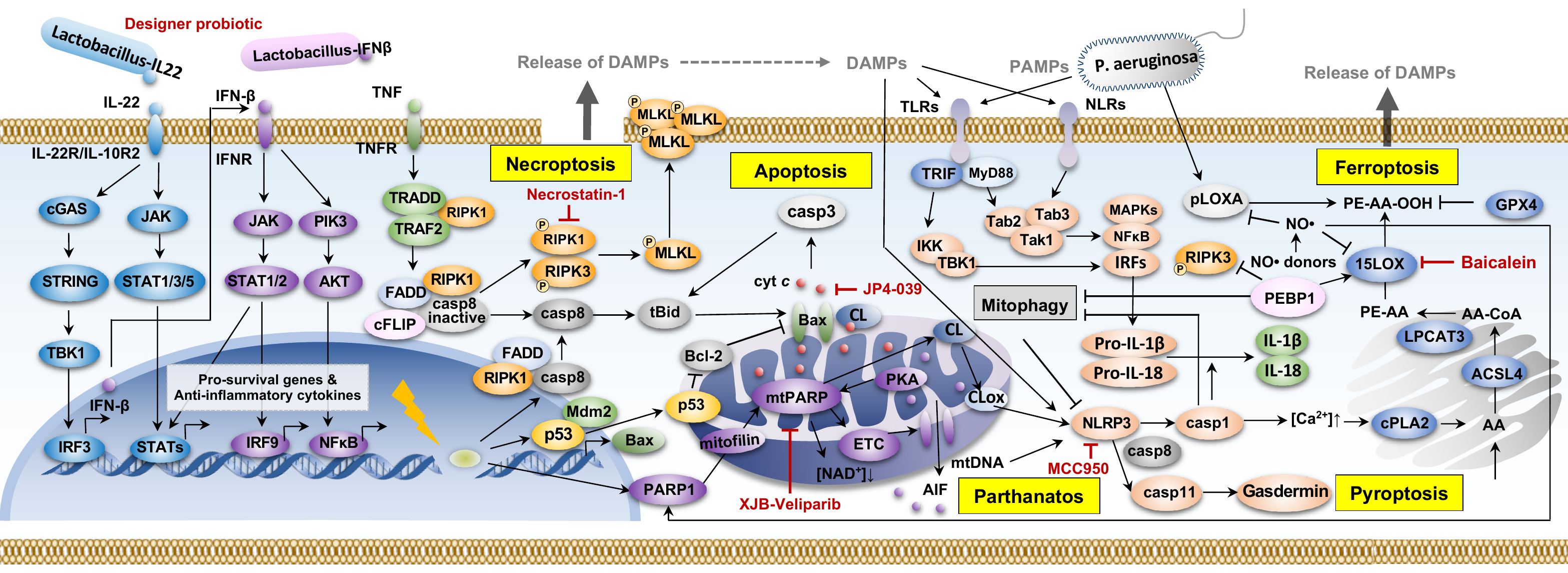}}
\vspace{-6pt}
\caption{Distinct cell death pathways induced by irradiation exposure.}
\vspace{-12pt}
\label{pathway}
\end{figure*}

We also developed {\em statistical model checking} (SMC) techniques to calibrate and analyze ODE systems with probabilistic initial states~\cite{Palaniappan2013,liu2019statistical,ramanathan2015parallelized}. In this setting, bounded linear temporal logic is used to encode quantitative behavioral constraints and qualitative properties of biochemical networks. By equipping existing parameter search algorithms with a SMC-based evaluation method, one can arrive both novel and efficient parameter estimation and analysis methods. This framework has been generalized to deal with stochastic rule-based models~\cite{liu2016parameter,wang2016formal} and hybrid automata \cite{gyori2015approximate} and applied to studies of various biological systems such as innate immune system~\cite{Liu2016} and cell death/survival pathways~\cite{Liu2014,liu2017quantitative,kagan2017oxidized,kapralov2020redox}.

However, in many settings, it is not fruitful to view the functioning of a biological system in terms of a single entity. Rather, the system will possess multiple operational {\em modes} with a specific signaling network being active in each mode. For example, signaling responses to ionizing irradiation exposure within individual cells follow distinct cell death pathways (Fig.~\ref{pathway}). The interconnectivity between these pathways brings up new challenges on determining the sequence and timing of medication administration against radiation injuries, since interventions may alter the cellular state and induce differential modes of dynamics~\cite{steinman2018improved,thermozier2019radioresistance,thermozier2020anti}. 

In such situations, using a monolithic approach will result in a messy and very large model with time-invariant parameters. Consequently, the modeling and analysis approach based on the notion of {\em multi-mode} biological networks will be very useful. Multi-mode dynamics and their modeling appear frequently in physical and engineering settings~\cite{goebel2009hybrid}. Multiple variants of the formalism called {\em hybrid automata} \cite{henzinger96} are often used to model biological processes \cite{tomlin04,ye08,aihara10,antoniotti03,lincoln04,baldazzi11}. However such models are difficult to analyze and to get around this, most efforts end up imposing severe restrictions on the dynamic laws associated with the modes~\cite{clarke2003verification,henzinger1999discrete,agrawal2006behavioural}.

\begin{figure}[h]
\centerline{\includegraphics[width=3.5in]{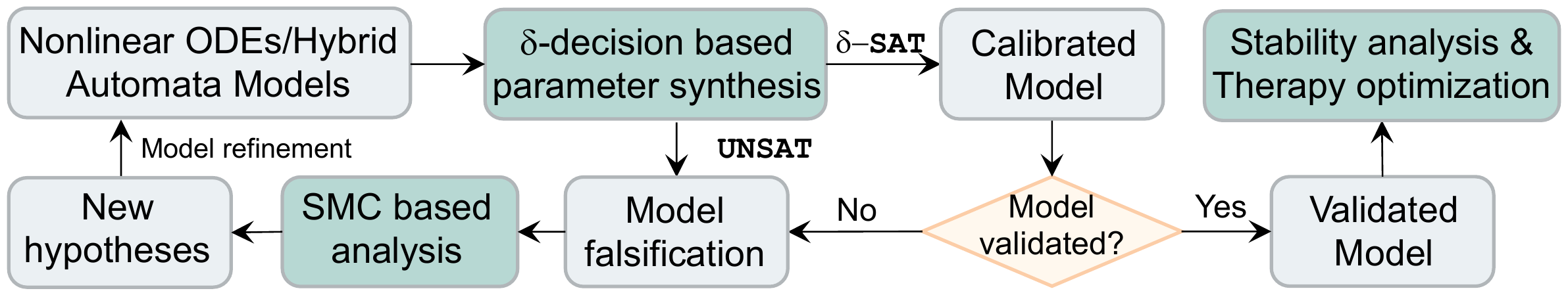}}
\caption{A model checking-enabled modeling framework.}
\vspace{-6pt}
\label{framework}
\end{figure}

In this paper, we present a novel model checking-based framework for analyzing single- or multi-mode biological systems with nonlinear dynamics (Fig.~\ref{framework}). Given a dynamical system, we describe the set of states of interest as a first-order logic formula and perform {\em bounded model checking} to determine reachability of these states. We adapt an interval constrains propagation (ICP) based algorithm to explore the parameter spaces and identify the sets of parameters using which the model satisfies desired behavior constraints. Note that determining the truth value of first-order sentences over the reals with nonlinear real functions is a well-known undefinable problem. We use $\delta$-decision procedures~\cite{gao12a} to ask for answers that may have one-sided $\delta$-bounded errors. If the model satisfies the desired behavior (e.g. the model matches training and testing data), we can carry out analysis tasks such as stability analysis or identify novel therapeutic strategies using $\delta$-decision procedures. Otherwise, we will conduct SMC-based analysis~\cite{liu2019statistical} to generate new hypotheses and refine the model structure iteratively. We demonstrate the applicability of our framework by ``proof-of-concept'' studies on prostate cancer and cardiac disorders~\cite{Liu2014b,Liu2015,liu2014parameter}.

Turning to related work, a survey of modeling and analysis of biological systems using hybrid models can be found in~\cite{luca08}. Formal verification of hybrid systems is a well-established domain~\cite{alur}. Analyzing the properties of biochemical networks using model checking techniques is being actively pursued by a number of groups~\cite{clarke08,chabrier04,kwiatkowska08,miyano11}. Of particular interest in our context are parameter synthesis methods which identify parameter values for which some qualitative behavior~\cite{rovergene,donze,khalid2018calibration}. 

The rest of the paper is organized as follows. We briefly introduce model checking in the next section. We describe the $\delta$-decision procedures in Section~\ref{delta}. We then present the applications of $\delta$-decision-based analysis methods in Section~\ref{analysis}. In the final section, we summarize and discuss future work.

\section{Model Checking}

Amir Pnueli introduced temporal logics into the world of program verification \cite{pnueli1977temporal}. The algorithmic verification procedure called {\em model checking} was formulated by Clarke and Emerson and independently by Joseph Sifakis \cite{clarke2008birth}. Briefly, the model checking procedure operates as follows. Given a model $\mathcal{M}$ with initial state $s$, a model checker decides if a property written as a temporal logic formula $\phi$ is satisfied, denoted as $\mathcal{M},s \models \phi$. This can be done by: (i) constructing a finite (state) transition system corresponding to $\mathcal{M}$ in which each state represents a possible configuration of the system and each transition represents an evolution of the system from one configuration to another, and (ii) verify whether $\phi$ is satisfied by exhaustively exploring the set of system executions. The key feature of the model checking procedure is that it is fully automated. Further, if it is not the case that $\mathcal{M},s \models \phi$ then the procedure will usually return as a ``counter-example'' an execution due to which the property is not being met by the system. This can serve as a powerful debugging tool for systems that are large and complex. An excellent starting point for exploring this whole field is \cite{clarkebook}.

\section{$\delta$-Decision Procedures}
\label{delta}

\subsection{$\lrf$-Formulas and $\delta$-Decisions Over The Reals}
In~\cite{gao12a}, we developed a theory of decision problems over the reals with computable functions. Most common continuous real functions are computable, including 
solutions of Lipschitz-continuous ODEs. 
In fact, the notion of computability of real functions directly corresponds to whether they can be numerically simulated. We write $\mathcal{F}$ to denote an arbitrary collection of symbols representing computable functions over $\mathbb{R}^n$ for various $n$. We consider the first-order formulas with a signature $\lrf = \langle 0,1,\mathcal{F},>\rangle$. Note that constants are seen as 0-ary functions in $\mathcal{F}$. $\lrf$-formulas are evaluated in the standard way over the corresponding structure $\mathbb{R}_{\mathcal{F}}= \langle \mathbb{R}, \mathcal{F}, >\rangle$.
We use atomic formulas of the form $t(x_1,...,x_n)>0$ or $t(x_1,...,x_n)\geq 0$, where $t(x_1,...,x_n)$ are built
up from functions in $\mathcal{F}$. To avoid extra preprocessing of formulas, we give an explicit definition of $\lrf$-formulas as follows.

\begin{definition}[$\lrf$-Formulas]
Let $\mathcal{F}$ be a collection of Type 2 functions, which contains at least
$0$, unary negation -, addition $+$, and absolute value $|\cdot|$. We define:
\begin{align*}
t& := x \; | \; f(t(\vec x)), \mbox{ where }f\in \mathcal{F}\mbox{, possibly
constant};\\
\varphi& := t(\vec x)> 0 \; | \; t(\vec x)\geq 0 \; | \; \varphi\wedge\varphi
\; | \; \varphi\vee\varphi \; | \; \exists x_i\varphi \; |\; \forall x_i\varphi.
\end{align*}
In this setting $\neg\varphi$ is regarded as an inductively defined operation
which replaces atomic formulas $t>0$ with $-t\geq 0$, atomic formulas $t\geq 0$
with $-t>0$, swaps $\wedge$ and $\vee$, and swaps $\forall$ and $\exists$.
Implication $\varphi_1\rightarrow\varphi_2$ is defined as
$\neg\varphi_1\vee\varphi_2$.
\end{definition}

\begin{definition}[Bounded Quantifiers]
We define
\begin{align*}
\exists^{[u,v]}x.\varphi &=_{df}\exists x. ( u \leq x \land x \leq v \wedge
\varphi),\\
\forall^{[u,v]}x.\varphi &=_{df} \forall x. ( (u \leq x \land x \leq v)
\rightarrow \varphi),
\end{align*}
where $u$ and $v$ denote $\lrf$ terms whose variables only
contain free variables in $\varphi$, excluding $x$. It is easy to check that
$\exists^{[u,v]}x. \varphi \leftrightarrow \neg \forall^{[u,v]}x. \neg\varphi$.
\end{definition}

We say a sentence is bounded if it only involves bounded quantifiers.

\begin{definition}[Bounded $\lrf$-Sentences]
A {\em bounded $\lrf$-sentence} is
$$Q_1^{[u_1,v_1]}x_1\cdots Q_n^{[u_n,v_n]}x_n\;\psi(x_1,...,x_n).$$
$Q_i^{[u_i,v_i]}$s are bounded quantifiers, and $\psi(x_1,...,x_n)$ is a
quantifier-free $\lrf$-formula.
\end{definition}

We write $\psi(x_1,...,x_n)$ as $\psi[t_1(\vec
x)>0,...,t_k(\vec x)>0; t_{k+1}(\vec x)\geq 0,...,t_m(\vec
x)\geq 0]$ to emphasize that $\psi(\vec x)$ is a Boolean
combination of the atomic formulas shown.

\begin{definition}[$\delta$-Variants]\label{variants}
Let $\delta\in \mathbb{Q}^+\cup\{0\}$, and $\varphi$ an
$\lrf$-formula of the form
$$\varphi: \ Q_1^{I_1}x_1\cdots Q_n^{I_n}x_n\;\psi[t_i(\vec x, \vec y)>0;
t_j(\vec x, \vec
y)\geq 0],$$ where $i\in\{1,...k\}$ and $j\in\{k+1,...,m\}$. The {\em
$\delta$-weakening} $\varphi^{\delta}$ of $\varphi$ is
defined as the result of replacing each atom $t_i > 0$ by $t_i >
-\delta$ and $t_j \geq 0$ by $t_j \geq -\delta$. That is,
$$\varphi^{\delta}:\ Q_1^{I_1}x_1\cdots Q_n^{I_n}x_n\;\psi[t_i(\vec x, \vec
y)>-\delta; t_j(\vec x,
\vec y)\geq -\delta].$$
\end{definition}

We then have the following main decidability result.
\begin{theorem}[$\delta$-Decidability]
Let $\delta\in\mathbb{Q}^+$ be arbitrary. There is an algorithm which, given any bounded $\varphi$, correctly returns one of the following two answers:
\begin{itemize}
	\item $\phi$ is false ($\mathsf{unsat}$);
	\item $\phi^\delta$ is true ($\delta$-$\mathsf{sat}$).
\end{itemize}
Note when the two cases overlap, either answer is correct.
\end{theorem}

We call this new decision problem the $\delta$-decision problem for
$\lrf$-sentences.
\begin{definition}[$\delta$-Complete Decision Procedures]
If an algorithm solves the $\delta$-decision problem correctly for a set $S$ of $\lrf$-sentences, we say it is $\delta$-complete for $S$.
\end{definition}

From $\delta$-decidability, $\delta$-complete decision procedures always exist for bounded $\lrf$-formulas. In practice, we have shown that the combination of the DPLL(T) framework and ICP indeed gives us a $\delta$-complete decision procedure~\cite{dreal}. 

\subsection{Parameterized $\lrf$-Representations of Hybrid Automata}

We now describe hybrid automata using $\lrf$-formulas, and define parameterization and perturbations on them.

A hybrid system is a tuple $H = \langle X$, $Q$, $\flow$, $\guard$, $\reset$, $\inv$, $\init\rangle$
where $X\subseteq \mathbb{R}^n$ specifies the range of the {\em continuous variables}  $\vec x$ of the system. $Q=\{q_0,...,q_m\}$ is a finite set of discrete {\em control modes}. $\flow \subseteq Q\times X\times \R \times X$ specifies the {\em continuous dynamics} for each mode. The $\flow$ predicate is usually defined either as explicit mappings from $\vec a_0$ and $t$ to $\vec a_t$,  or as solutions of systems of differential equations/inclusions that specify the derivative of $\vec x$ over time. $\jump\subseteq Q\times X\times Q\times X$ specifies the {\em jump conditions} between modes. $\inv \subseteq Q\times X$ defines the {\em invariant conditions} for the system to stay in a control mode. $\init \subseteq Q\times X$ defines the set of {\em initial configurations} of the system. Without loss of generality we always assume that $q_0$ is the only intial mode, and $\init_{q_0}\subseteq X$ denotes the initial values for the continuous variables.
\begin{definition}[$\lrf$-Representations]\label{lrf-definition}
\index{$\lrf$-Representation}
Let $H = \langle X$, $Q$, $\flow$, $\jump$, $\inv$, $\init\rangle$ be an $n$-dimensional hybrid automaton.  Let $\mathcal{F}$
be a set of real functions, and $\mathcal{L}_{\mathbb{R}_{\mathcal{F}}}$ the corresponding first-order language. We say that $H$ has an $\lrf$-representation, if for every $q,q'\in Q$, there exists  quantifier-free
$\lrf$-formulas $$\phi^q_{\flow}(\vec x, \vec x_0, t), \phi^{q\rightarrow
q'}_{\jump}(\vec x,
\vec x'), \phi^{q}_{\inv}(\vec x), \phi^q_{\init}(\vec x)$$
such that for all
$\vec a ,\vec a'\in \mathbb{R}^n$,
$t\in\mathbb{R}$:
\begin{itemize}
\item $\mathbb{R}\models \phi^q_{\flow}(\vec a, \vec a', t)$ iff $(q, \vec
a,
\vec a', t)\in \flow$.
\item $\mathbb{R}\models \phi^{q\rightarrow q'}_{\jump}(\vec a, \vec a')$ iff
$(q, q', \vec a, \vec a')\in \jump$.
\item $\mathbb{R}\models \phi^q_{\inv}(\vec a)$ iff $(q, \vec a)\in \inv.$
\item $\mathbb{R}\models \phi^q_{\init}(\vec a)$ iff $q = q_0$ and $\vec a\in
\init_{q_0}$.
\end{itemize}
\end{definition}
We can write $H = \langle X, Q, \phi_{\flow}, \phi_{\jump}, \phi_{\inv},\phi_{\init}\rangle$ to emphasize that $H$ is $\lrf$-represented. But from now on we simply write $\flow, \jump, \inv, \init$ to denote these logic formulas, so that we can use $H = \langle X, Q, \flow, \jump, \inv, \init\rangle$ directly to denote the $\lrf$-representation of $H$.

\begin{definition}[Computable Representation]
We say a hybrid automaton $H$ has a {\em computable representation}, if $H$ has
an $\lrf$-representation, where $\mathcal{F}$ is an arbitrary set of computable
functions.
\end{definition}

Combining continuous and discrete behaviors, the trajectories of hybrid systems are {\em piecewise continuous}. This motivates a two-dimensional structure of time, with which we can keep track of both the discrete changes and the duration of each continuous flow.
\begin{definition}[Hybrid Time Domain]
A {\em hybrid time domain} $T$ is a subset of $\mathbb{N}\times \mathbb{R}$ of the form
$T_m=\{(i, t): i<m \mbox{ and } t\in [t_i, t_i']\mbox{ or }[t_i, +\infty)\},$
where $m\in \mathbb{N}\cup\{+\infty\}$, $\{t_i\}_{i=0}^m$ is an increasing sequence in $\mathbb{R}^+$, $t_0= 0$, and $t_i'=t_{i+1}$.
\end{definition}

We write the set of all hybrid time domains as $\mathbb{H}$.

\begin{definition}[Hybrid Trajectories]
Suppose $X\subseteq\mathbb{R}^n$ and $T_m$ is a hybrid time domain. A {\em hybrid trajectory} is any continuous function $\xi: T_m\rightarrow X.$
\end{definition}

We write $\Xi_X$ to denote the set of all possible hybrid trajectories from $\mathbb{H}$ to $X$.
We can now define trajectories of a given hybrid automaton. The intuition behind the following definition is straightforward. The labeling function $\sigma_{\xi}^H(i)$ is used to map a step $i$ to the corresponding discrete mode in $H$. In each mode, the system flows continuously following the dynamics defined by $\flow(q, \vec x_0, t)$. Note that $(t-t_k)$ is the actual duration in the $k$-th mode. When a switch between two modes is performed, it is required
that $\xi(k+1, t_{k+1})$ is updated from the exit value $\xi(k, t_k')$ in the previous mode, following the jump conditions.

\begin{definition}[Trajectories of a Hybrid Automaton]\label{trajec}
Let $H$ be a hybrid automaton, and $\xi: T_m\rightarrow X$ a hybrid trajectory.
We say that $\xi: T_m\rightarrow X$ is {\em a trajectory of $H$ of discrete depth $m$}, if there
exists a {\em labeling function} $\sigma^H_{\xi}: \mathbb{N}\rightarrow Q$ such
that:
\begin{itemize}
\item $\sigma^H_{\xi}(0) = q_0$ and
$\mathbb{R}_{\mathcal{F}}\models \init_{q_0}(\xi(0,0))$.
\item For any $(i, t)\in T_m$,
$\mathbb{R}_{\mathcal{F}}\models \inv_{\sigma^H_{\xi}(i)} (\xi(i,t))$.
\item When $i=0$, $\mathbb{R}_{\mathcal{F}}\models\flow_{q_0}(\xi(0,0), \xi(0,t), t)$.
\item When $i = k+1$, where $0< k+1 <m$,
\begin{eqnarray*}
\mathbb{R}_{\mathcal{F}}&\models&\flow_{\sigma^H_{\xi}(k+1)}(\xi(k+1, t_{k+1}),\xi(k+1, t), \\
&&(t - t_{k+1}))\mbox{ and }\\
\mathbb{R}_{\mathcal{F}}&\models& \jump_{(\sigma^H(k)\rightarrow
\sigma^H(k+1))}(\xi(k, t_k'), \xi(k+1,t_{k+1})).
\end{eqnarray*}
\end{itemize}
\end{definition}
We write $\llbracket H\rrbracket$ to denote all possible trajectories of $H$.

\begin{definition}[Reachability Properties]
Let $H$ be a hybrid automaton and $U\subseteq X\times Q$ be a subset of its state space.
Let $U\subseteq X\times Q$ be a subset of the state space of $H$. $H$ reaches $U$ if there exists $\xi\in \llbracket H\rrbracket$ such that there exists $t\in \mathbb{R}$ and $n\in\mathbb{N}$ satisfying
$$(\xi(t,n), \sigma_{\xi}^{H}(n))\in U.$$
\end{definition}
Let $H$ be a hybrid system.
Parameter synthesis for reachability properties asks for a set of parameters such that some mode can be reached.

\begin{definition}[Parameterized Hybrid Automaton]
We say a hybrid automaton $H$ is parameterized by $\vec p$, if
We say $H$ is parameterized by $\vec p = (p_1,...,p_m)$, if
$$H(\vec p) = \langle X, Q, \flow(\vec p), \jump(\vec p), \inv(\vec p), \init(\vec p)\rangle,$$
where $\vec p$ are among the free variables in the $\lrf$-representation of $H$.
\end{definition}
\begin{definition}[Parameter Synthesis for Reachability Properties]
Let $H(\vec p)$ be a hybrid automaton parameterized by variables $\vec p = (p_1,...,p_m)$, and $U\subseteq X\times Q$ a subset of its state space.
Thus, the parameter synthesis problem for reachability asks for an assignment for $\vec a\in \mathbb{R}^m$ such that $H(\vec a)$ reaches $U$.
\end{definition}

\subsection{Synthesizing Parameters with $\delta$-Decisions}

We now show how to encode parameter synthesis problems for $\lrf$-represented hybrid systems using $\lrf$-formulas. 
Throughout the following two definitions, let $H = \langle X$, $Q$, $\flow$, $\jump$, $\init\rangle$ be an $n$-dimensional $\lrf$-represented hybrid system with $|Q|=m$, and $\unsafe$ an $\lrf$-formula that encodes a subset $U\subseteq X\times Q$. Let $k\in \mathbb{N}$ and $M\in \mathbb{R}$ be the bounds on steps and time respectively. Recall that $q_0\in Q$ always denotes the starting mode.

$\reach_{H,q'}^k(\vec x_k^t)$ defines the states that $H$ can reach, if after $k$ steps of discrete changes it is in mode $q'$. From there, if $H$ makes a $\jump$ from mode $q'$ to $q$, then the states have to make a discrete change following $\jump_{q'\rightarrow q}(\vec x_k^t, \vec
x_{k+1})$. As last, in mode $q'$, any state $\vec x_{k+1}^t$ that $H$ can reach
should satisfy the $\flow$ conditions $\flow_q(\vec x_{k+1}^t, \vec x_{k+1}, t)$
in mode $q$. 
Note that after each discrete jump, a new time variable $t_k$ is
introduced and independent from the previous ones.

The $(k,M)$-reachability encoding of $H$ and $U$, $\reach^{k,M}(H,U)$, is defined as:
\begin{flalign*}
&\exists \vec a \exists^X \vec x_0 \exists^X\vec x_0^t\cdots \exists^X\vec
x_k\exists^X \vec x_k^t \exists^{[0,M]}t_0\cdots \exists^{[0,M]}t_k&\\
&\Big(\ \init_{q_0}(\vec x_0)\wedge \flow_{q_0}(\vec a, \vec x_0, \vec x_0^t,
t_0)&\\
&\wedge \forall^{[0,t_0]}t\forall^X\vec x\;(\flow_{q_0}(\vec a, \vec x_0, \vec x,
t)\rightarrow \inv_{q_0}(\vec a, \vec x))&\\
&\wedge
\bigvee_{i=0}^{k-1}\Big( \bigvee_{q, q'\in Q} \Big(\jump_{q\rightarrow q'}(\vec a, \vec
x_i^t, \vec x_{i+1})\wedge \flow_{q'}(\vec a, \vec x_{i+1}, \vec
x_{i+1}^t, t_{i+1})&\\
&\wedge \forall^{[0,t_0]}t\forall^X\vec x\;(\flow_{q'}(\vec a, \vec x_{i+1}, \vec x,
t)\rightarrow \inv_{q_0}(\vec a, \vec x)) )\Big)\Big)&\\
&\wedge\ \unsafe(\vec a, \vec x_{k}^t)\Big).&
\end{flalign*}
$H$ reaches $U$ in $k$ steps of discrete jumps with time duration less than $M$ for each state iff $\reach^{k,M}(H,U)$ is true.

\section{Applications}
\label{analysis}

\subsection{Model Calibration and Falsification}
The $\delta$-decision problems can be solved using our dReal tool~\cite{dreal}. Parameter estimation of single-mode ODE models can be encoded as SMT formulas by BioPSy~\cite{madsen2015biopsy} and solved by dReal, while for multi-mode models we ask a $k$-step reachability question: Is there a set of parameter values using which the model reaches the goal region in $k$ steps? The dReach tool~\cite{dreach} can automatically build such reachability formulas from a multi-mode model and a goal description, which are then verified by dReal. If $\mathsf{unsat}$ is returned, the model is {\em unfeasible}, which means that the model is unable to satisfy a desired behavior no matter which parameter values are used. This can be used to reject model hypotheses. For example, we have showed that the Fenton-Karma model~\cite{fk} of cardiac cells is unable to reproduce the ``spike-and-dome'' morphology of action potential which has been observed in epicardial cells~\cite{Liu2014b}. On the other hand, if the model is $\delta$-$\mathsf{sat}$, a witness (i.e. a set of parameter values) is returned.
For example, using the Bueno-Cherry-Fenton model~\cite{bcf}, we have identified critical parameter ranges that can cause cardiac disorders such as tachycardia and fibrillation~\cite{Liu2014b}.

\subsection{Identification of Therapeutic Strategies}

\begin{figure}[htbp]
\centerline{\includegraphics[width=3.5in]{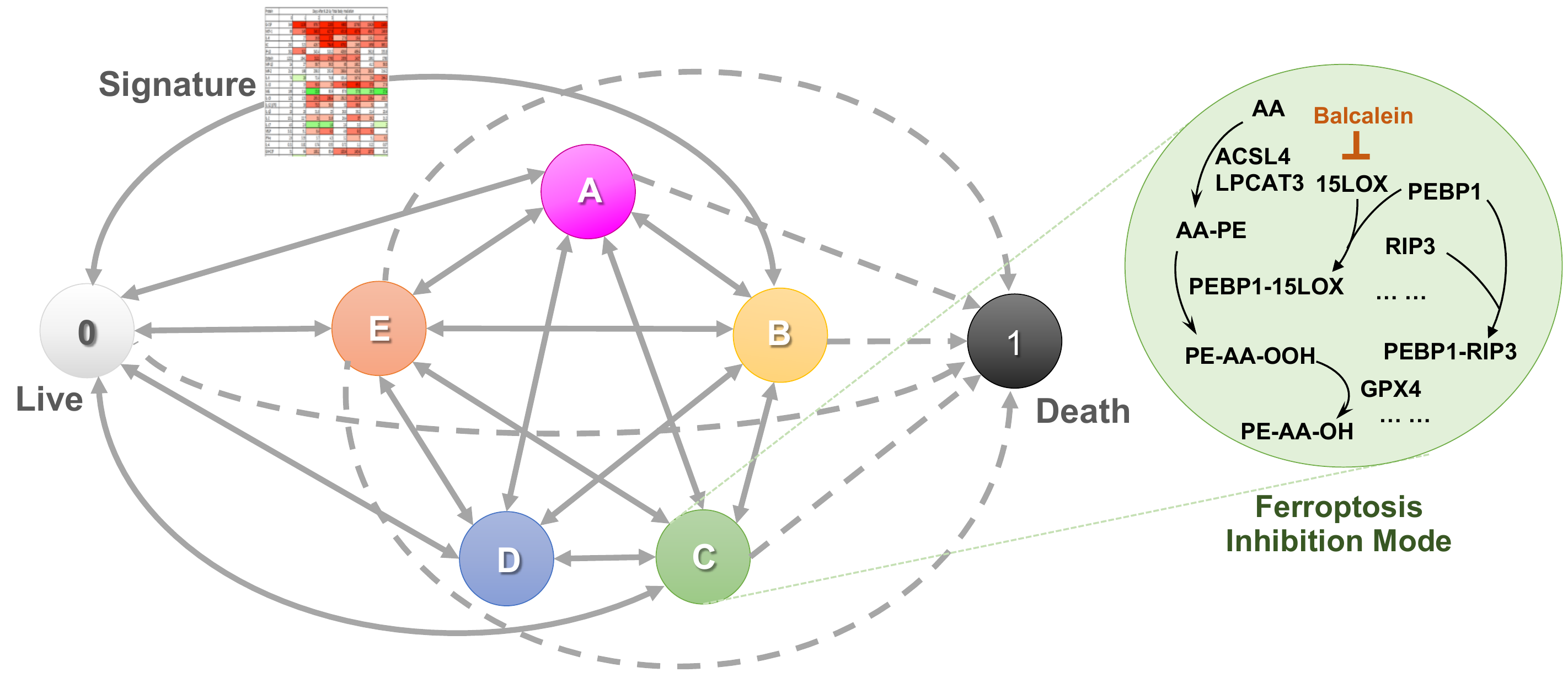}}
\caption{Multi-mode modeling. Mode 0: live cell, no treatment; Mode A: apoptosis inhibition; Mode B: necroptosis inhibition; Mode C: Ferroptosis inhibition; Mode D: Pyroptosis inhibition; Mode E: Parthanatos inhibition Mode 1: dead cell.}
\vspace{-6pt}
\label{scheme}
\end{figure}

The $\delta$-decision procedures can be used to design optimal therapeutic strategies. For example, Fig.~\ref{scheme} illustrates a multi-mode model of the TBI-induced signaling network shown in Fig.~\ref{pathway}. The starting point Mode 0 corresponds to live cells under no treatment for 24 h after TBI. Mode 1 is the ``point of no return'' which leads to cell death. Modes A-E represent live cells subjected to particular treatments (e.g. Balcalein-induced ferroptosis inhibition). The jump conditions are defined by the molecular signature. For example, starting with Mode 0, if the oxidized CL level exceeds a threshold $\theta_1$, the system jumps to Mode A, which means that we deliver the apoptosis inhibitor JP4-039. The system then evolves according to the ODEs in Mode A, until another jump condition--e.g. the level of activated RIP3 above threshold $\theta_2$--is reached, which implies the onset of necroptosis, or Mode B. We then deliver necroptosis inhibitor necrostatin-1. Suppose the system next jumps back to Mode 0 and stays. The mode path $0 \rightarrow A \rightarrow B  \rightarrow 0$ suggests a successful treatment scheme defined by a set of jump conditions. Note that we also aim to minimize the number of drugs used (i.e. path length) to avoid potential side effects. Thus, the problem of determining which drug to deliver at what time, evolves into a parameter synthesis problem for hybrid automata and can be tackled by $\delta$-decision procedures. In a ``proof-of-concept'' study~\cite{Liu2015}, we have used this approach to identity personalized therapeutic strategies for prostate cancer patients.

\subsection{Stability Analysis} 
The {\em robustness} of biological systems often refers to the consistency of system behavior in response to small perturbations. For example, cardiac cells filter out insignificant stimulations to ensure proper functioning in noisy environments. Using the $\delta$-decision procedures, we can verify this by checking if the action potential can be successfully triggered by a small range of stimulation. An $\mathsf{unsat}$ answer returned by dReach will guarantee that the model is robust to the corresponding stimulation amplitude~\cite{Liu2014b}.  

In addition to ``time-bounded'' robustness, the $\delta$-decision procedures can also be used to analyze the infinite-time stability of nonlinear dynamical systems~\cite{kong2018,gao2019}. Such stability--often referred as {\em structural stability}--exists in the systems that have unique globally asymptotically stable steady states. A standard way of verifying this property is to find a {\em Lyapunov function}~\cite{sontag} that provides theoretical guarantees on qualitative behavior of the system. Lyapunov-enable analysis has been applied to mass-action law based kinetic models of signaling networks such as T-cell kinetic proofreading and ERK signaling~\cite{al2020computational}. Our $\delta$-decision procedures enable the Lyapunov stable analysis for systems with non-polynomial nonlinearity in two ways: (i) Given a template function, we can synthesize a Lyapunov function by solving $\exists\forall$-formulas that encode the corresponding non-convex, multi-objective and disjunctive optimization problem~\cite{kong2018}; (ii) We can provide a sound and relative-complete proof system for induction rules that robustify the standard notions of Lyapunov functions~\cite{gao2019}.

\section{Conclusion and Future Works}
We have presented a model checking-enabled framework for analyzing systems biology models with the help of $\delta$-decision procedures. We used the $\lrf$-formulas to describe parameterized hybrid automata and encode parameter synthesis problems. We employed the $\delta$-decision procedures to perform bounded model checking and obtain parameters using which the model can satisfy desired properties. We have showed that the $\delta$-decision procedures can be used to estimate unknown model parameters, reject model hypothesis, analyze the model's Lyapunov stability, as well as design combination therapies. These tasks can be assembled as a unified workflow for understanding the mechanism of diseases and identifying therapeutic options. Our preliminary studies on prostate cancer, cardiac disorders, and radiation diseases have shown promising results. 

An interesting direction is extending our method for probabilistic settings to address the inherent variability in biological systems. To cope with the model complexity, an idea is to approximate the hybrid system as a multi-mode network of DBNs by extending the approximation technique we have developed for a single system of ODEs~\cite{Liu2012b}. We plan to explore this in our future work. In this respect, the Chow-Liu tree representation-based approximation scheme used in~\cite{pichene2019modeling} promises to offer helpful pointers.

\section*{Acknowledgment}

This work is supported by National Institutes of Health grants P01DK096990, U19AI068021, and P41GM103712.

\bibliographystyle{IEEEtran}
\bibliography{IEEEabrv,ref}

\end{document}